\documentclass[twocolumn, aps, prb, floatfix]{revtex4}
\usepackage{graphicx}
\newcommand{\+}[1]{#1^\dagger}
\newcommand{\xop}[1]{#1^\times}

\begin{document}
\title{Non-Markovian entanglement dynamics in the presence of system-bath coherence}
\author{Arend~G.~Dijkstra}
\author{Yoshitaka Tanimura}
\affiliation{Department of Chemistry, Graduate School of Science, Kyoto University, Kyoto 606-8502, Japan}
\begin{abstract}
  A complete treatment of the entanglement of two-level systems, which evolves through the contact with a thermal bath, must include the fact that the system and the bath are not fully separable. Therefore, quantum coherent superpositions of system and bath states, which are almost never fully included in theoretical models, are invariably present when an entangled state is prepared experimentally. We show their importance
  for the time evolution of the entanglement of two qubits coupled to independent baths. In addition, our treatment is able to handle slow and low-temperature thermal baths.
\end{abstract}
\date\today
\maketitle

The crucial feature of quantum information is the presence of coherent superpositions. A single isolated two-level system (qubit) can be prepared in a superposition of its 0 and 1 states, and the manipulation of such states
leads to new possibilities for the storage and processing of information. In a pure quantum system, the
superposition is entirely coherent, which means that a definite phase relation exists between the 0 and 1 states. 
Unlike in the ideal isolated case, the interactions of real quantum systems with their environment lead to the destruction of these phase relations, in other words, decoherence. 

More interesting than a single two-level system is a collection of multiple such qubits. Coherent superposition states of multiple qubits can be prepared, and their dynamics are studied to understand the functioning of quantum networks.\cite{Cao.2009.jcpa.113.13825} The presence of phase relations between qubits,
a second type of coherence, is termed entanglement. The destruction of entanglement through interactions with the environment is an important problem, both for a fundamental understanding of quantum mechanics and the development of quantum information processing.\cite{Yu.2009.science.323.598} Recently, it has been found that the decay of entanglement can be very different from that of the single qubit coherence. In particular, the entanglement can disappear completely in finite time.\cite{Yu.2004.prl.93.140404} Interactions between qubits further affect the time scale of the decay.\cite{Ban.2009.pra.80.032114}

Here, we argue that, besides single qubit coherence and the entanglement of multiple qubits, there is a third form of coherence that is important for the description of quantum information. The physical nature of this effect lies in a detailed description of the heat bath that leads to decoherence and disentanglement. In a complete theory, both the system
and the bath are quantum mechanical. It is therefore possible, and often unavoidable in experiment, to create coherent superpositions of system and bath states, and such superpositions
will play an important role in the time evolution of the qubit system of interest. 

The description of entanglement dynamics starts from an equation of motion for the qubits together with an initial condition, which are both affected by a correct description of the bath. Most theoretical treatments, including the Redfield and Lindblad formulations, describe the qubit dynamics under the Born and ultrafast bath approximations.\cite{footnote1} These lead to convenient time-local equations of motion, but are only valid if the time scale of bath dynamics is much faster than the characteristic time scales of the system.\cite{BPbook} Furthermore, in the same limit, the initial state can be chosen to be of a form in which system and bath are independent. If the fast bath condition is not fulfilled, however, the approximation breaks down. As briefly explained below, it is also invalid at low temperature, where quantum fluctuations contribute longer time scales.\cite{Tanimura.2006.jpsj.75.082001} 

Although the effect of noise correlation (or non-Markovian dynamics) on the entanglement has been studied under the rotating wave approximation at zero temperature,\cite{Bellomo.2009.arxiv09100050v1} the role of the bath on the initial condition has rarely been investigated. However, in the regime where non-Markovian effects are important, the presence of system-bath correlations invalidates the initial state in which the system and the bath are independent. Especially in the experimentally relevant case where the qubit system is excited out of equilibrium and the subsequent dynamics is probed, a proper treatment of the initial state is crucial.

It is the role of the bath in the equation of motion and, in particular, in the initial state, that we explore in this paper. We present a new method to calculate the entanglement
that rigorously deals with the quantum dynamics in this situation. The method treats the system-bath interaction non-perturbatively and without assuming a fast noise bath.\cite{Tanimura.2006.jpsj.75.082001, Tanimura.1989.jpsj.58.101, Ishizaki.2005.jpsj.74.3131, Tanaka.2009.jpsj.78.073802} It thereby enables us to study the role of system-bath coherence and of an initial state in which a coherent superposition of the qubit system and the bath is prepared.

Before introducing our rigorous approach, we clarify the role of system-bath coherence by starting from a simple model, in which both the system and the bath are two-level systems with transition energy $\epsilon$. In this much simplified situation, the time scales of system and bath evolution are identical, and therefore the role of system-bath coherence is expected to be strong. With a coupling 
$g$ the complete Hamiltonian is given by $H = \epsilon \+{c}_S c_S + \epsilon \+{c}_B c_B + g (\+{c}_S c_B + \+{c}_B c_S)$,
where $c_{S/B}$ and $\+{c}_{S/B}$ are the usual annihilation and creation operators working on the system or bath. The system plus bath are assumed to be in thermal equilibrium at an inverse temperature $\beta$.
Apart from a shift in the effective excitation energy, no effect of the bath is apparent in the reduced density matrix for the system degrees of freedom $\rho$. By considering the full density matrix $R$ of system plus bath, however, one finds a matrix element that represents coherence between the system and the bath, $\langle 0_S 1_B | R | 1_S 0_B \rangle =  -\frac{1}{Z} e^{-\beta \epsilon} \sinh \beta g$,
where $S$ denotes the system and $B$ the bath, and $Z$ is the total partition function. This term would be zero for the commonly considered initial condition $R = \rho \otimes \exp(-\beta H_B) / \mathrm{tr} \exp(-\beta H_B)$. Its presence shows that the density matrix in equilibrium is not separable; it cannot be written as the product of a system and a bath part. The correct treatment of the system-bath coherence in the initial state is important for the entanglement dynamics following a pulse that excites the system out of thermal equilibrium.

In the following, we will study the entanglement of two qubits each separately coupled to a more general bath, which evolves on a characteristic time scale $1/\gamma$.
The qubits are labeled $1$ and $2$, both have an excitation energy $\epsilon$, and they are coupled by an interaction $J$. 
The system Hamiltonian is
\begin{equation}
  H_\mathrm{S} = \epsilon (\+{c}_1 c_1 + \+{c}_2 c_2) + J (\+{c}_1 + c_1)(\+{c}_2 + c_2).
\end{equation} 
To study system-bath coherence, we need to introduce a fully quantum-mechanical bath. A bath that is sufficiently general to model many physical systems, while at the same time allowing efficient calculations, is given by a set of harmonic oscillators.\cite{Leggett.1987.rmp.59.1} The individual bath modes are labeled with an index $j$, and have masses $m_j$, frequencies $\omega_j$, coordinates $x_j$ and corresponding momenta $p_j$. The Hamiltonian for the system-bath interaction plus the harmonic bath is ($\hbar = 1$ throughout the paper)
\begin{equation}
  H_\mathrm{SB+B} = - \sum_{\alpha, j} g_{\alpha j} V_\alpha x_j + 
     \sum_j \left(\frac{p_j^2}{2 m_j} + \frac{m_j \omega_j^2 x_j^2}{2}\right).
\end{equation}
Starting from this Hamiltonian, we need to derive an equation of motion for the reduced density matrix of the system. It must describe the fluctuations in the system energies and the dissipation arising from the interaction with the bath. Furthermore, we want the method to be able to describe an initial state in which the system and bath are correlated, i.e. for which the complete density matrix cannot be written as a direct product of a system and a bath part. 

In the reduced description, all necessary information on the bath and the system-bath coupling is contained in the spectral densities $\mathcal J_{\alpha\alpha'}(\omega) = \sum_j \frac{g_{\alpha j} g_{\alpha'j}}{2 m_j \omega_j} \delta(\omega - \omega_j)$.
While the $V_\alpha$'s can be any system operator, we focus here on two independent baths which induce transitions in a single qubit each, given by $V_\alpha = c_\alpha + \+{c}_\alpha$ for $\alpha = 1, 2$. 
Thus, the spectral representations of the cross-correlation functions are identically zero, $\mathcal J_{12}(\omega) = \mathcal J_{21}(\omega) = 0$. 
Furthermore, we assume that each bath evolves on the same time scale and couples to a qubit with the same strength. Both baths can then be described with the same spectral density function, $\mathcal J_{11}(\omega) = \mathcal J_{22}(\omega) =: \mathcal J(\omega)$, which we model as $\mathcal J(\omega) = \omega  \frac{2 \lambda \gamma}{\gamma^2 + \omega^2}$. 
To see the effect of fluctuations in the system parameters and of the dissipation originating from this spectral density, we consider the correlation function of the noise variables. It can be calculated as the Fourier transform of the spectral density to be
\begin{equation}
  L(t) = \sum_{k=0}^\infty c_k e^{-\nu_k |t|},
\end{equation}
where we have defined the time scales $1/\nu_{k} = \beta / 2 \pi k$ (for $k \geq 1$) and $1/\nu_{0} = 1/\gamma$ and the prefactors $c_{0} = \lambda \gamma(-i + \cot \beta \gamma/2)$, $c_{k} = (4 \lambda \gamma / \beta) \nu_{k} / (\nu_{k}^2 - \gamma^2)$. 
For large $\gamma$, that is, in the case of a fast bath, the dissipative term, given by the imaginary part of the correlation function, becomes delta correlated. Quantum effects, however, make the fluctuations induced in the system (given by the real part of $L(t)$) always correlated for low temperature, in which case the evolution of the bath contains time scales fixed by the Matsubara frequencies $\nu_k$.\cite{Tanimura.2006.jpsj.75.082001} The fast bath (Markovian) approximation is then invalid.
Note that we do not make the rotating wave approximation in the system-bath interaction, which can lead to significant changes in entanglement dynamics.\cite{Yang.2009.cpb.18.4662}

Using the Feynman-Vernon influence functional approach, the reduced equation of motion can be written in the form of path integrals over the system coordinates. The path integrals can be evaluated numerically,\cite{Makri.1998.jpca.102.4414, Mak.1991.pra.44.2352}
or the equation of motion can be rewritten as a stochastic Schr\"odinger equation.\cite{Stockburger.2002.prl.88.170407} Alternatively, a time-local equation of motion for the reduced density matrix can be derived.\cite{Tanimura.1989.jpsj.58.101,  Ishizaki.2005.jpsj.74.3131, Tanaka.2009.jpsj.78.073802} It accounts for the dynamics of the quantum bath and the system-bath coherence through a set of auxiliary density matrices, whose time evolution is also governed by time-local equations of motion. The set of density matrices can be propagated efficiently in a computer, and this description is therefore well-suited for the study of entanglement dynamics.
The complete equation of motion involves multiple auxiliary density matrices, labeled by a multi-index $n_{\alpha k}$, where $\alpha$ labels the various terms in the system-bath interaction, and $k$ the Matsubara frequencies included to allow for low temperature. It includes the time scales and prefactors from the correlation function, and is given by\cite{Ishizaki.2005.jpsj.74.3131, Tanimura.2006.jpsj.75.082001}
\begin{eqnarray}
  \dot \rho^n(t) &=& - \left( i \xop H_S + \sum_{\alpha=1}^2 \sum_{k=0}^M n_{\alpha k} \nu_{k} \right) \rho^n(t) \\
   &-& \sum_{\alpha=1}^2 \left( \frac{2\lambda}{\beta \gamma} - i \lambda - \sum_{k=0}^M \frac{c_{k}}{\nu_{k}} \right) \xop V_\alpha \xop V_\alpha \rho^n(t) \nonumber \\
   &-& i \sum_{\alpha=1}^2 \sum_{k=0}^M \xop V_\alpha \rho^{n_{\alpha k} \to n_{\alpha k}+1}(t) \nonumber \\
   &-& i \sum_{\alpha=1}^2 \sum_{k=0}^M n_{\alpha k}\left(c_{k} V_\alpha \rho^{n_{\alpha k}^-}(t) - c_{k}^* \rho^{n_{\alpha k}^-}(t) V_\alpha \right), \nonumber
\end{eqnarray}
where $\xop A B \equiv [A, B]$.
The notation $n_{\alpha k} \to n_{\alpha k}+1$ refers to an increase in the $\alpha k$'th component of the multi-index, while all other indices are unchanged. Similarly, $\rho^{n_{\alpha k}^-} = \rho^{n_{\alpha k} \to n_{\alpha k} -1}$ denotes a decrease of this index. For the simulations in this paper, we will use the values of $\epsilon = 1.5 J$, $\lambda = 0.3 J$, $\gamma = 0.5 J$, $\beta = 2.5 J$. This method has recently been applied to calculate the exciton dynamics in light-harvesting complexes.\cite{Ishizaki.2009.jcp.130.234111, Strumpfer.2009.jcp.131.225101, Chen.2009.jcp.131.094502}

Given the equation of motion, the dynamics of the system can be calculated numerically starting from an arbitrary initial condition. The procedure takes the system-bath coupling into account without invoking perturbative, Markovian, or rotating wave approximations. The system-bath coherence in the initial state can be included through the auxiliary density matrices. In particular, to obtain a thermal equilibrium initial state, which will in general be a partially coherent superposition of the system and the bath, one can propagate the equation of motion starting from any state for a time longer than any characteristic time scale in the system or the bath. Here, we will not consider the possibility of a non-ergodic system, in which case additional averaging is required. We will now first discuss the effect of the bath in the equation of motion, followed by a treatment of the correlated initial state.

\begin{figure}[t]
 \includegraphics{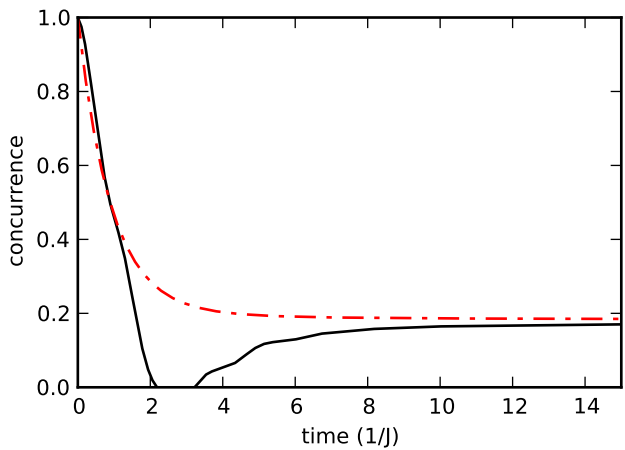}
\caption{\label{fig:toeq} Decay of entanglement from an initially maximally entangled state $|\phi\rangle = (|1\rangle - |2\rangle)/\sqrt{2}$. The solid line is the result from our full calculation while the Redfield result is plotted as a dash-dotted line for comparison.}
\end{figure}

A popular measure of the entanglement of two qubits is the entanglement of formation, which can be calculated directly from the reduced density matrix through the concurrence.\cite{Wootters.1998.prl.80.2245}
The effect of slow bath dynamics has been shown to fundamentally alter the time evolution of entanglement between two quantum systems at zero temperature. In particular, strong coupling to a bath can not only change the time scale on which entanglement disappears, but also lead to revival of entanglement after a period of zero concurrence.\cite{Bellomo.2007.prl.99.160502} This effect is also observed in the current model. In Fig.\ref{fig:toeq}, the concurrence is plotted as a function of time, starting from an initially entangled state $|\phi\rangle = (|1\rangle - |2\rangle)/\sqrt{2}$, and the ad-hoc but common assumption of an independent bath. The dash-dotted line shows the result of a Redfield calculation in the secular approximation, which treats the system-bath interaction perturbatively and in the Born and Markov approximations.\cite{BPbook, Ishizaki.2009.jcp.130.234111} Without the secular approximation, the Redfield equation can lead to an unphysical density matrix, and the concurrence is ill-defined. The result from the full calculation (shown as a solid line) is markedly different from the Redfield predictions. Before reaching the same positive equilibrium concurrence as in the Redfield calculation, it shows a revival of concurrence (at $t \approx 3$) after sudden death of entanglement (around $t \approx 2$). This calculation shows that conventional quantum master equations do not completely describe the entanglement dynamics. To understand the positive equilibrium value of the concurrence, which is the combined effect of the coupling between the two qubits and the temperature, we consider the following simplified model. 

For two qubits weakly interacting with a heat bath, one would expect the reduced density matrix in equilibrium to be given by the Gibbs measure $\rho_\mathrm{eq} = e^{-\beta H_S} / \mathrm{Tr} e^{-\beta H_S}$.
For a system Hamiltonian $H_S = \epsilon ( \+{c}_1 c_1 + \+{c}_2 c_2 ) + J (\+{c}_1 c_2 + \+{c}_2 c_1)$,
the concurrence in this equilibrium state is
\begin{equation}
  C = \frac{-1 + \sinh \beta J}{\cosh \beta \epsilon + \cosh \beta J},
\end{equation}
which is positive for $\beta J > \sinh^{-1} 1 \approx 0.88$. We thus find that for large $J$ or at low temperatures, the equilibrium concurrence is positive. 
This suggests that it might be better to consider the entanglement of the eigenstates of the system Hamiltonian instead, which vanishes in an equilibrium state obtained through weak coupling with a heat bath.

This is only half of the story. Apart from its effect on the equation of motion, the slow bath also makes it impossible to factorize the initial state into a system and a bath part. This becomes clear if we consider a realistic physical initial state. In the previous discussion, the initial state was assumed to be a product state of system and bath density matrices. This is non-physical in the case of strong system-bath coupling, because it is usually hard, if not impossible, to prepare such a state. 

\begin{figure}[t]
 \includegraphics{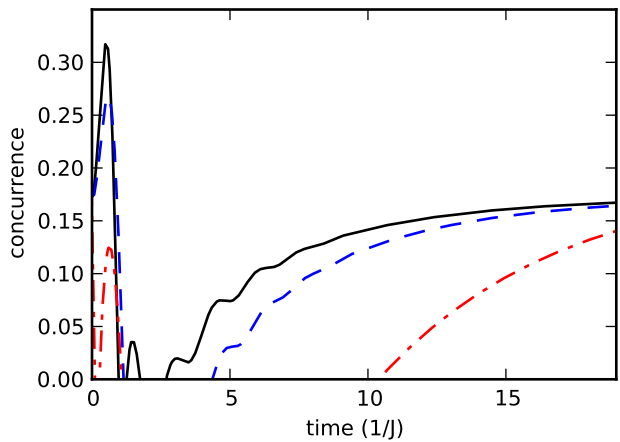}
\caption{\label{fig:concurrence} Time evolution of the entanglement after a pulse has been applied to the equilibrium state. The solid line shows the result of the full calculation, while the system-bath coherence at $t=0$ was neglected in the results presented as a dashed line. The dash-dotted line shows the Redfield result.}
\end{figure}

A more realistic initial state can be obtained by applying a pulse to the equilibrium state. Here, we consider a pulse that affects only the first qubit by rotating it over an angle $\pi$. This operation can be expressed as $\rho \to \sigma_1^y \rho \sigma_1^{y*}$. Such an operation clearly takes the system out of equilibrium, and it creates additional entanglement. It can be realized for example as a $\pi$-pulse in nuclear magnetic resonance.\cite{Levitt.book} The resulting time evolution of the concurrence is shown in Fig.~\ref{fig:concurrence} as a solid line. The entanglement dynamics is not captured in the Redfield approach (which cannot include the system bath correlations in the initial state), shown as a dash-dotted line. To see the effect of system-bath coherence explicitly, we also perform the calculation for a factorized initial state (resulting in the dynamics plotted as a dashed line in Fig.~\ref{fig:concurrence}). Although the reduced density matrices for both the system and the bath are the same as in the full calculation, the system-bath coherence was set to zero just before applying the spin flip operation at $t=0$. Even though the system-bath coherence is fully accounted for in the subsequent time evolution in both cases, it is clear that the presence of the system-bath coherence in the initial state changes the entanglement dynamics dramatically. The initially created entanglement is much larger than in the case of a product initial state. Moreover, the effect lasts long enough to reduce the entanglement death time considerably. Death of entanglement occurs only for short periods of time in the full calculation, while no entanglement is found until after $t=4$ when the initial system-bath correlation is ignored. This finding shows that, even in calculating a property such as the entanglement that depends only on the system degrees of freedom, the initial system-bath coherence must be properly taken into account. In effect, memory of entanglement can be stored in the bath. While we have uncorrelated baths coupled to each qubit in this study, the effect of correlation in bath modes might well further strengthen this finding, and entanglement measures that include the system-bath coherence will be needed.

In conclusion, we have studied the entanglement of two qubits in the presence of a quantum mechanical bath. The coherence between system and bath is found to have an important effect on the time evolution of the concurrence when the system is excited out of equilibrium, which is not correctly described by a conventional quantum master equation. These findings will be relevant for the design of quantum networks\cite{Cao.2009.jcpa.113.13825} and information devices, as well as for the dynamics of excitations in biological systems,\cite{Ishizaki.2009.jcp.130.234111, Strumpfer.2009.jcp.131.225101, Dijkstra.2010.njp} quantum dots,\cite{Kubota.2009.jpsj.78.114603} and conjugated polymers.\cite{Dykstra.2009.jpcb.113.656} For spin systems, the effect of a fermionic bath should be investigated.\cite{Khaetskii.2002.prl.18.186802}

We thank dr. A Ishizaki for helpful comments on the manuscript.

\end{document}